# Estimates for the thermodynamic signatures of vortex-lattice melting in conventional superconductors

O. Bossen and A. Schilling
Physik-Institut der Universität Zürich,
Winterthurerstrasse 190, 8057 Zürich, Switzerland

**Abstract**

The first-order nature of the vortex-lattice melting transition in copper-based layered high-$T_c$ superconductors is well established. The associated discontinuities in magnetization have been extensively studied, for example, in $YBa_2Cu_3O_7$ [1, 2] and $Bi_2Sr_2CaCu_2O_8$, [3] while the respective latent heats have been systematically investigated only in $YBa_2Cu_3O_7$ and related compounds [4, 5, 6, 7, 8, 9, 10, 11, 12, 13]. The apparent absence of such signatures in conventional superconductors such as Nb raises the question whether or not the concept of vortex-lattice melting is applicable at all in such materials [14]. Based on available literature to describe the vortex-state and using the Lindemann criterion, we estimate quantitatively the order of magnitude for the expected latent heats of melting and the associated discontinuities in magnetization, respectively, as functions of a few known material parameters. It turns out that both thermodynamic quantities are not strictly vanishing even in isotropic materials as long as $\kappa > 1/\sqrt{2}$, but they are small and may often be beyond the available experimental resolution.

## 1. Introduction

In 1957 Abrikosov wrote his seminal work [15] on two distinctly different types of superconductors. He predicted that in so-called type II superconductors, an external magnetic field penetrates a superconductor in form of magnetic flux lines which arrange themselves in a regular lattice, each one carrying one magnetic flux quantum $\varphi_0$ = 2.07x10$^{-15}$ Vs. It was conjectured later that this lattice might undergo some kind of first-order "melting" transition [16, 17] and turn into a "liquid" at sufficiently high temperatures, but well below the transition to superconductivity. This prediction was subsequently invoked to explain the sudden onset of damping of mechanical oscillators bearing a superconducting sample in a magnetic field [18, 19], hysteresis in the resistivity [20] and neutron-scattering data [21]. The theoretical properties of vortex matter were discussed exhaustively [22, 23, 24], but the existence of a true first-order transition has been of speculative nature until the first strong thermodynamic evidence for it was found in magnetization measurements on $Bi_2Sr_2CaCu_2O_8$ [3, 25]. The measurement of the associated latent heat through specific-heat measurements on $YBa_2Cu_3O_7$ followed soon after [10, 4], and thermodynamic consistency with magnetization data [1] was demonstrated [2,4]. It became clear that high-quality crystals are necessary in order to clearly observe the first-order behaviour, because defects such as twinning boundaries have been shown to suppress the transition and render it to second order [10], presumably because they lead to a glass-like phase [26]. Many experimental reports in the literature indeed describe second-order or glass-like transitions, the exact nature of which is influenced by the details of vortex pinning in a given sample. Although the nature of the high-temperature ("liquid") phase is essentially unexplored and no direct experimental proofs for the existence of vortices as distinct entities in this phase are available, we will call the phase transition under discussion hereafter "melting transition".

All confirmed measurements of first-order melting transitions have been made on layered cuprates [1, 2, 3, 4, 5, 6, 7, 8, 10, 11, 25, 27, 28], and they are now theoretically fairly well understood [17, 29, 30]. Following the same theoretical arguments, a temperature-driven phase transition of the vortex lattice should, in principle, also occur in sufficiently clean low-temperature superconductors. However, there are only very few reports on measurements of thermodynamic quantities in the context of vortex-lattice melting in such compounds. For Nb and $Nb_3Sn$, for example, attempts have been made to measure the melting entropy directly [14, 31], and the absence of a related signal in very pure Nb has led to reflections about the complete absence of vortex-lattice melting in this compound [14]. $Nb_3Sn$ seems to be the only low $T_c$ superconductor with a report of first-order like features in thermodynamic quantities [31], which still awaits an independent experimental confirmation, however.

In order to quantitatively understand such experimental results we derive here explicit estimates for the expected discontinuities in entropy and in magnetization for various type-II superconductors. These estimates are based on established theoretical work in the literature about the location of the conjectured vortex-lattice melting lines in the magnetic phase diagram, on Richard's rule [32] to estimate the melting entropy per particle, on the use of the correct "single-vortex length" [33], and on taking the enhancement of the resulting configurational entropy by the strong temperature dependence of relevant model parameters near the upper critical field into account [29, 30]. We conclude that a measurement of the melting entropy and a related discontinuity in the magnetization, if they exist in conventional superconductors, should be feasible on selected compounds using state-of the art techniques provided that vortex pinning is weak enough.

## 2. Basic melting theory

*a) The melting lines*

The assumptions of the simplest theory about the melting of the vortex lattice are very similar to a basic melting theory for solids. In 1910, Lindemann introduced a melting criterion [34], based on the idea that melting occurs as soon as the thermal mean-square displacements $\langle u_{th}^2 \rangle^{1/2}$ of the atoms in a lattice reaches a certain fraction of the lattice constant $a$, i.e., $\langle u_{th}^2 \rangle^{1/2} \approx c_L a$ with the Lindemann number $c_L < 1$. This heuristic argument has proved to be reliable, although the value for $c_L$ may vary widely for different crystal systems. Corresponding quantitative calculations for the vortex-lattice melting lines $B_m(T)$ (by comparing a calculated $\langle u_{th}^2 \rangle^{1/2}$ for vortices with the mean vortex distance $a_0 = \sqrt{\varphi_0 / B}$) lead to an implicit equation [17, 35]

$$\frac{t_m}{\sqrt{1-t_m^2}} = \frac{2\pi c_L^2}{Gi^{1/2} b^{1/2} f(b)}, \qquad (1)$$

with $t_m = T_m / T_{c0}$ the reduced melting temperature $T_m$ for $B = B_m$ with respect to the critical temperature $T_{c0}$ in zero magnetic field, $b = B / B_{c2}(T)$ the reduced magnetic induction $B$ with respect to the temperature dependent upper-critical field $B_{c2}(T)$, and $f(b)$ a model-dependent function (see below). The Ginzburg number $Gi$ is a measure for the width of the fluctuation region in zero magnetic field around $T_{c0}$. We consider here a uniaxial case in which the external magnetic field is applied along the crystal direction with the smallest upper critical field ("*c*-axis"). In SI-units, $Gi$ is then defined as

$$Gi = \frac{1}{2}\left(\frac{\mu_0 \mathrm{k_B} T_{c0}}{4\pi B_c(0)^2 \varepsilon \xi(0)^3}\right)^2 = \frac{\pi \mu_0^2 \mathrm{k_B^2} T_c^2 \kappa^4}{B_{c2\,GL}(0)\varepsilon^2 \phi_0^3}, \qquad (2)$$

where $\mu_0 = 4\pi \times 10^{-7}$ Vs/Am is the permeability of the vacuum, $\mathrm{k_B} = 1.38\times 10^{-23}$ J/K is the Boltzmann constant, $\kappa$ is the Ginzburg-Landau parameter for $B//c$, $\varepsilon < 1$ is the anisotropy parameter, here defined as the ratio of $B_{c2}$ in the *c*-direction and in a direction perpendicular to it, $\xi$ is the coherence length relevant for $B$//c (i.e., the „in-plane coherence length"), and $B_c(0)$ and $B_{c2GL}(0)$ are the thermodynamic and the upper-critical fields within the Ginzburg-Landau theory, respectively, linearly extrapolated from $T_{c0}$ to $T = 0$ K. The assumption of a linear $B_{c2}(T)$ does not account for experimentally determined values of $B_{c2}(0)$ that are often given in the literature, however, and it will not lead to an accurate estimate of the melting lines over the full range of temperatures. Therefore we will assume in the following a relationship of the form

$$B_{c2}(T) = B_{c2}(0)\left(1 - \left(T/T_{c0}\right)^2\right) \qquad (3)$$

with $B_{c2GL}(0) = 2B_{c2}(0)$. The empiric formula (3) can be made asymptotically correct for low and high fields by modifying the parameter $B_{c2}(0)$. Any deviation from the simple Eq. (3) in a given material can be accounted for, if necessary, by replacing the term $1 - t_m^2$ in Eq. (1) by an appropriate formula [17, 35].

The function $f(b)$ in Eq. (1) has been first calculated by Houghton, Pelcovits and Sudbø [17] for $\kappa >> 1$,

$$f_{HPS}(b) = \left(\frac{1.657}{\sqrt{1-b}} + 1\right) \Big/ (1-b) \ .$$

(4)

In their original work, a linear $B_{c2}(T)$ was assumed, which we have replaced in Eq. (1) by Eq. (3). Mikitik and Brandt calculated $f(b)$, again for $\kappa >> 1$, within a collective pinning theory [35] to

$$f_{MB}(b) = \frac{2\beta_A}{1-b} \frac{\sqrt{1+(1+c(b))^2} - 1}{c(b)(1+c(b))}, \qquad (5)$$

valid for all values of $b$ throughout the mixed state, with $c(b) = \sqrt{\beta_A(1-b)}/2$ and $\beta_A = 1.16$ the Abrikosov number. The functions $f(b)$ from Eqs. (4) and (5) only slightly differ between the two approaches. The quantities $f(b)(1-b)^{3/2}$ vary only slowly with $b$ between ≈ 2.66 for $b$ = 0 and ≈ 1.73 for $b$ = 1.

In order to be able to briefly discuss also the important case of an arbitrary $\kappa$, it is instructive to consider a simplified version of Eq. (1),

$$\mathrm{k_B} T_m = A c_{66} {c_L}^2 {a_0}^3 \varepsilon, \qquad (6)$$

with the shear modulus $c_{66}$. To interpret Eq. (6) we consider the result of Brandt [36] within a theory of weak collective pinning,

$$c_{66} \approx \frac{B_{c2}(T)^2}{\mu_0} \frac{b(1-b)^2}{8\kappa^2} \left(1 - \frac{1}{2\kappa^2}\right)\left(1 - 0.58b + 0.29b^2\right), \qquad (7)$$

which was computed for *all* values of $\kappa > 1/\sqrt{2}$ and $b$ throughout the mixed state. The Eq. (6) together with Eq. (7) asymptotically coincides in the limit $b \to 0$ with the Houghton, Pelcovits and Sudbø formula (Eqs. 1 and 4) for $A$ ≈ 15 and with that of Mikitik and Brandt (Eq. 5) for $A$ ≈ 17, and in the limit $b \to 1$ within the same approaches for $A$ ≈ 33 and $A$ ≈ 31, respectively, if the Volume $a_0^3\varepsilon$ in Eq. (6) is replaced in this limit by $V_0$ from Eq. (14), see below.

The Eqs. (6) and (7) show in a very transparent way that for a given $T_m$ and $\kappa \to 1/\sqrt{2}$ (i.e., when approaching the type-I limit with $c_{66} \to 0$), the magnetic-flux density $B_m(T)$ approaches zero because $a_0 \to \infty$. In terms of the external magnetic field *H*, the melting field $H_m(T)$ aligns with the lower-critical field $H_{c1}(T)$ over a wide range of temperatures. In an alternative interpretation of this limit, a $\kappa \to 1/\sqrt{2}$ leads to a decrease of the Lindemann number to an effective value $\tilde{c}_L = c_L(1-1/2\kappa^2)^{1/2}$, or, in terms of the Ginzburg number in Eq. (1), to an increase of *Gi* to an effective $\tilde{G}i = Gi(1-1/2\kappa^2)^{-2}$.

In the following we will nevertheless use the melting lines obtained by Mikitik and Brandt [35] for $\kappa >> 1$ because they are explicitly valid for all values of *b* and *Gi* and also allow for a temperature dependence of the upper-critical field according to Eq. (3). We choose $c_L$ = 0.20 [37], and we will ignore for the moment any possible renormalization of *Gi* for the limit $\kappa \to 1/\sqrt{2}$, which would only affect Nb and $CaC_6$ listed in table 1.

*b) Distance of $B_m$ from the fluctuation region around $B_{c2}$*

In Fig. 1 we illustrate that besides $c_L$ it is mainly the Ginzburg number *Gi* (which can vary in different superconductors by orders of magnitude, see table 1) that determines the distance of the melting line from $B_{c2}(T)$. For most conventional superconductors to be discussed below *Gi* is small and the melting lines will therefore be close to the upper-critical field. However, even in isotropic superconductors *Gi* can be large enough so that a vortex-lattice melting transition distinct from $B_{c2}(T)$ might be observed in principle. For this case we have to additionally address the question whether $B_m(T)$ is located outside the critical-fluctuation region around $B_{c2}(T)$ or not. For the case $b \to 1$, which is of interest here, Mikitik and Brandt [35] have estimated the difference

$$\delta B = B_{c2}(T) - B_m(T) \approx \left(\frac{1.78}{2\pi c_L^2}\right)^{2/3} B_{c2}(0) t^{2/3} Gi^{1/3} (1-t^2)^{2/3}. \qquad (8)$$

The width of the critical-fluctuation region in a magnetic field on the temperature scale, on the other hand, has been estimated to [38]

$$\delta T_{fluct} \approx T_{c2}(B) Gi^{1/3} \left(\frac{B}{B_{c2}(0)}\right)^{2/3}, \qquad (9)$$

where $T_{c2}(B)$ is the inverted $B_{c2}(T)$. The corresponding width in $B$ around $B_{c2}(T)$ is

$$\delta B_{fluct} \approx \left| \frac{dB_{c2}}{dT} \right| \delta T_{fluct} = 2B_{c2}(0) t^2 Gi^{1/3} (1 - t^2)^{2/3}, \qquad (10)$$

where we have again made use of Eq. (3). The melting line is outside the critical region if $\delta B > \delta B_{fluct}$, or

$$t < 0.361/c_{L}, \qquad (11)$$

which is always fulfilled for reasonable values of $c_L$, *notably independently of the value of Gi*. A tighter condition, $\delta B > n\delta B_{fluct}$ with $n > 1$, will modify this criterion to $t < 0.361 n^{-3/4} / c_{L}$, but we may note that a clear first-order transition has been observed in $YBa_2Cu_3O_7$ up to $t \approx 0.98$ [7].

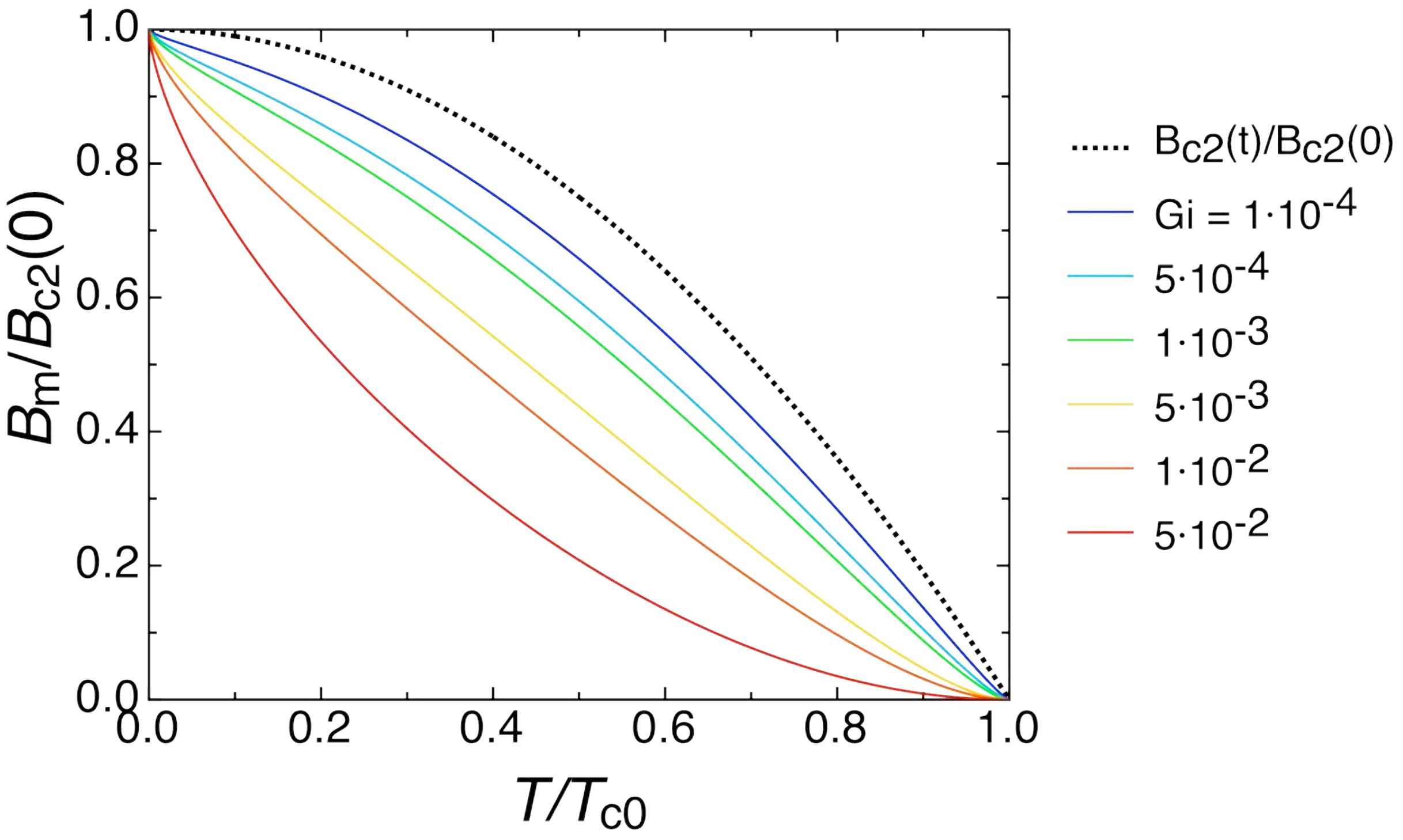

Fig. 1: Melting lines (according to Ref. [35] with $c_L = 0.20$) for different values of *Gi*.

*c) Discontinuities in entropy and in magnetization*

To obtain the melting entropies, we make use of Richard's rule [32] for crystal lattices, in which the configurational melting entropy per particle is assumed to be a constant multiple (or fraction) of $k_B$. With the volume $V_0$ occupied by one particle we then have

$$\Delta s_0 V_0 = \eta k_B, \tag{12}$$

where $\Delta s_0$ is the configurational melting entropy per volume and $\eta$ is an unknown and yet to be determined constant (see below).

To obtain the elementary volume $V_0$ for vortices that is relevant for counting the total number of degrees of freedom in the system, it is essential to use the correct "single-vortex length" $L_0$. This length has often been erroneously taken as the zero-temperature coherence length $\xi$ [14, 31], thereby vastly underestimating it and overestimating $\Delta s_0$. Kierfeld and Vinokur [33] calculated $L_0$ to

$$L_0 = \frac{\varepsilon\, a_0}{\sqrt{1-b}} \tag{13}$$

(which is much larger than $\xi$), and the volume $V_0$ becomes

$$V_0 = \frac{\varepsilon\, a_0^{\;3}}{\sqrt{1-b}} \tag{14}$$

which diverges as $B$ approaches the upper-critical field and therefore leads to a substantial reduction of $\Delta s_0$ in materials in which the melting lines are located very close to $B_{c2}(T)$.

It turned out that the entropy obtained from applying Richard's rule alone considerably underestimates the measured entropy changes $\Delta s$ upon vortex-lattice melting both in $YBa_2Cu_3O_7$ and in $Bi_2Sr_2CaCu_2O_8$ [29, 30]. Taking the marked temperature dependence of thermodynamic quantities into account, Dodgson *et al.* arrive at an enhancement of $\Delta s_0$ of the form [30]

$$\Delta s = \frac{\left[1-\tilde{b}+(2\tilde{b}-t^2)t^2\right]}{(1-t^2-\tilde{b})(1-t^2)}\Delta s_0, \tag{15}$$

with $\tilde{b} = B/B_{c2}(0)$. Although calculated in the London approximation, Eq. (15) contains corrections to account for the suppression of the order parameter around the vortex cores in the high-field limit $b \rightarrow 1$. This enhancement of $\Delta s_0$ may be particularly relevant for superconductors with small *Gi* where $B_{c2}(T) - B_m(T)$ is expected to be small, because it diverges at $B_{c2}(T)$ as $(1-b)^{-1}$ and therefore overcompensates the reduction of $\Delta s_0$ due to the diverging elementary vortex volume $V_0$, and also partly moderates the effect of the reduction of $c_{66}$ on $\Delta s_0$ as $B_m(T)$ approaches $B_{c2}(T)$, see Eqs. (14) and (7). While the $B_m(T)$ line of $YBa_2Cu_3O_7$ itself is, to some extent, also modified by the arguments raised in Ref. [30] because it is located very far from $B_{c2}(T)$ in this particular compound, the $B_m(T)$ in conventional superconductors with small *Gi* that are under discussion here must remain very close to $B_{c2}(T)$. However, we state that even small errors in estimating $B_m(T)$ may result in substantial uncertainties in $\Delta s$ (and therefore in $\Delta M$) because these quantities sensitively depend on $(1-b)$. In this sense, the corresponding quantities calculated in the remainder of this paper have to be taken as order-of-magnitude estimates, rather than as exact results.

In the context of vortex-lattice melting, the constant $\eta$ in Eq. (12) has been estimated to ≈ 0.16 [29] for $YBa_2Cu_3O_7$. To be consistent within our formalism, we have recalculated $B_m(T)$ and $\Delta$. for this compound according to the above rules (Eqs. 1, 5, 12 and 15) with the material parameters from table 1 and $c_L = 0.20$, and we obtain the best fit to the $\Delta s$ data from Ref. [7] for $H // c$ with $\eta = 0.077$ (see Fig. 2). To further justify this rather small value, we can use a crude estimate taken to explain $\Delta s$ in $Bi_2Sr_2CaCu_2O_8$ in Ref. 3, $T_m \Delta s = c_{66} c_L^2$, which, in combination with Eqs. (12) and (14), yields the correct melting line of Eq. (6) with $\eta$ = 1/*A*. For $b \to 0$, we then have $\eta \approx 0.06$ and for $b \to 1$, $\eta \approx 0.03$. It is possible, however, that $\eta$ assumes entirely different values in materials with lower *Gi*, and we therefore do not further specify the value of $\eta$ in our calculations.

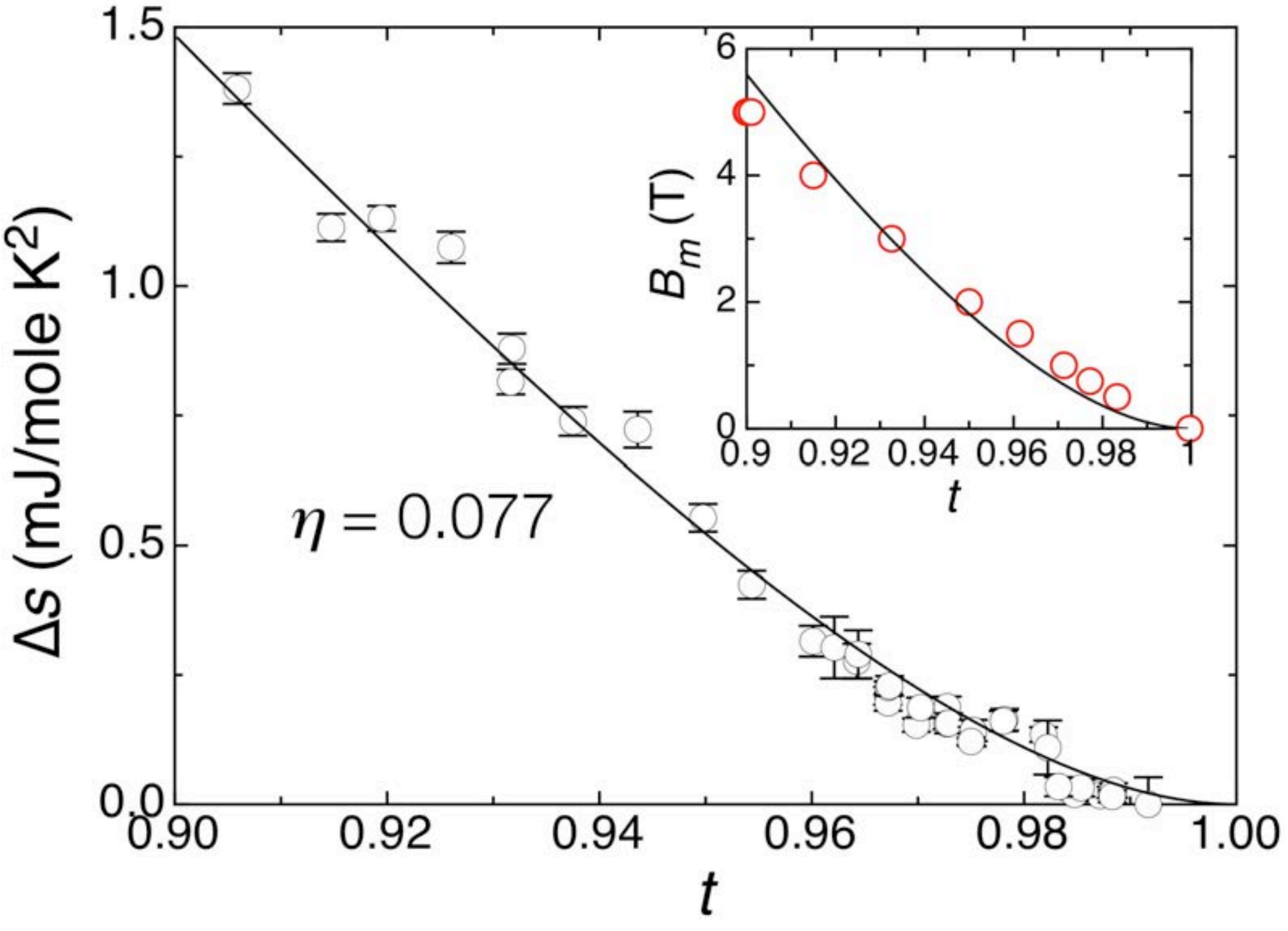

Fig. 2: Melting entropy $\Delta s$ and melting line $B_m(T)$ of $YBa_2Cu_3O_7$, calculated from the parameters given in table 1 and Eqs. 1, 5, 12 and 15. The $\Delta s$-data fit best for $\eta = 0.077$.

The resulting discontinuities in magnetizations can finally be derived from the Clausius-Clapeyron equation,

$$\Delta s = -\Delta M \frac{\mathrm{d}B_{\mathrm{m}}}{\mathrm{d}T} \tag{16}$$

with $\Delta M > 0$ , i.e., *M* increases when crossing $B_{\mathrm{m}}(T)$ from the solid to the "liquid" phase. It is interesting to note that in the limit $\kappa \to 1/\sqrt{2}$ (i.e., when approaching the type-I limit), $c_{66}$ in Eq. (7) vanishes even if the vortex density remains high in large magnetic fields, and so must the discontinuities in *s* and in *M* which are related to this energy scale. However, as long as $\kappa > 1/\sqrt{2}$, both $\Delta s$ and $\Delta M$ remain finite even in *isotropic* superconductors with $\varepsilon = 1$ because $B_{\mathrm{m}}(T)$ seems to stay outside the fluctuation region defined by Eq. (9) around $B_{\mathrm{c2}}(T)$.

## 3. Application to real materials

In order to estimate the order of magnitude of the discontinuities in entropy $\Delta s$ and in magnetization $\Delta M$ in real superconductors, we have compiled literature values of relevant material parameters for a number of superconductors of contemporary interest (see table 1). We have then calculated the melting lines $B_{\mathrm{m}}(T)$ according to Eqs. (1) and (5) with $c_{\mathrm{L}}$ = 0.20 (see Fig. 3).

In the Figs. 4-7 we have plotted the expected values for $\Delta s$ and $\Delta M$ (where we tentatively took the enhancement from Eq. (15) into account, but ignoring a possible influence on $B_{\mathrm{m}}$ and $\Delta s$ for the limit $\kappa \to 1/\sqrt{2}$), together with $\delta T$ and $\delta B$, the distances of the melting lines from $B_{\mathrm{c2}}(T)$ in *T* and in *B*, respectively. In Fig. 8, we also show a corresponding calculation for $\Delta s$ vs. *T* without the factor from Eq. (15) for comparison with Fig. 4a, to illustrate the impact of this correction on the order of magnitude of $\Delta s$.

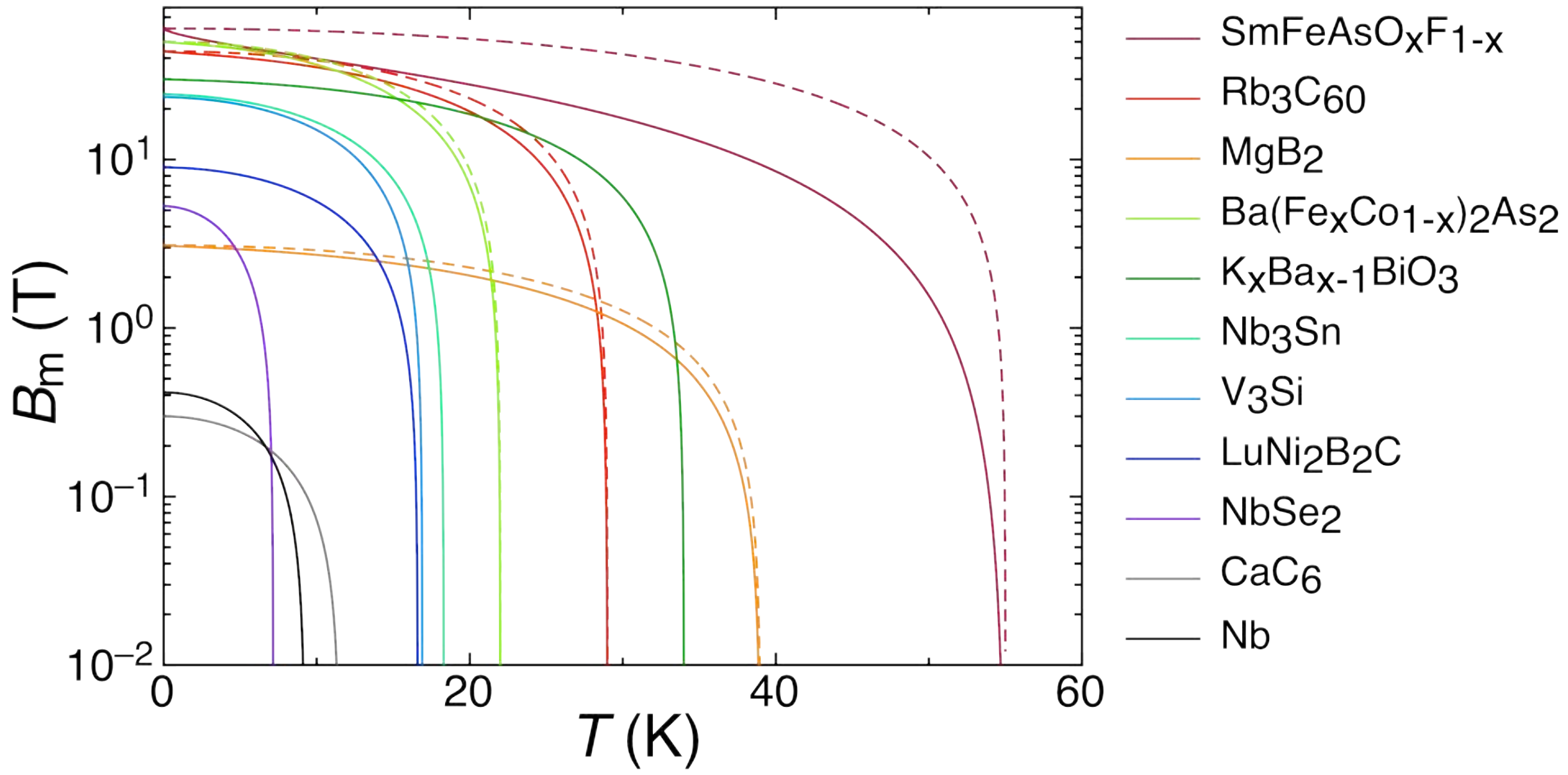

Fig. 3: Expected melting lines $B_m(T)$ for various type-II superconductors, calculated from Eqs. (1) and (5) with $c_L$ = 0.20 and the material parameters given in table 1 (solid lines). The $B_{c2}(T)$-lines (dotted lines) have been plotted only for $SmFeAsO_xF_{1-x}$, $Rb_3C_{60}$, $MgB_2$, and $Ba(Fe_xCo_{1-x})_2As_2$ for which they appear distinct from $B_m(T)$ in this representation.

a)

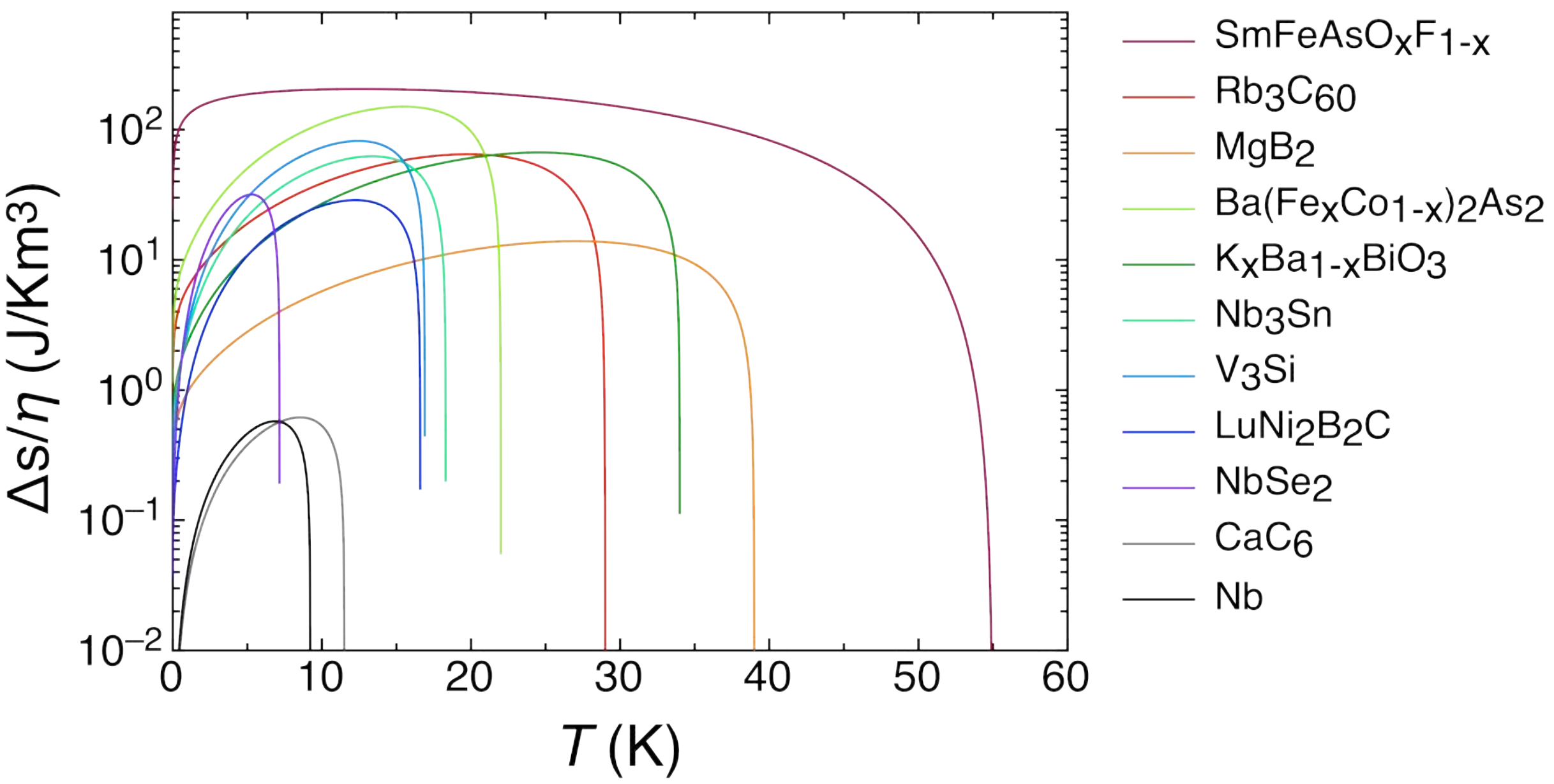

b)

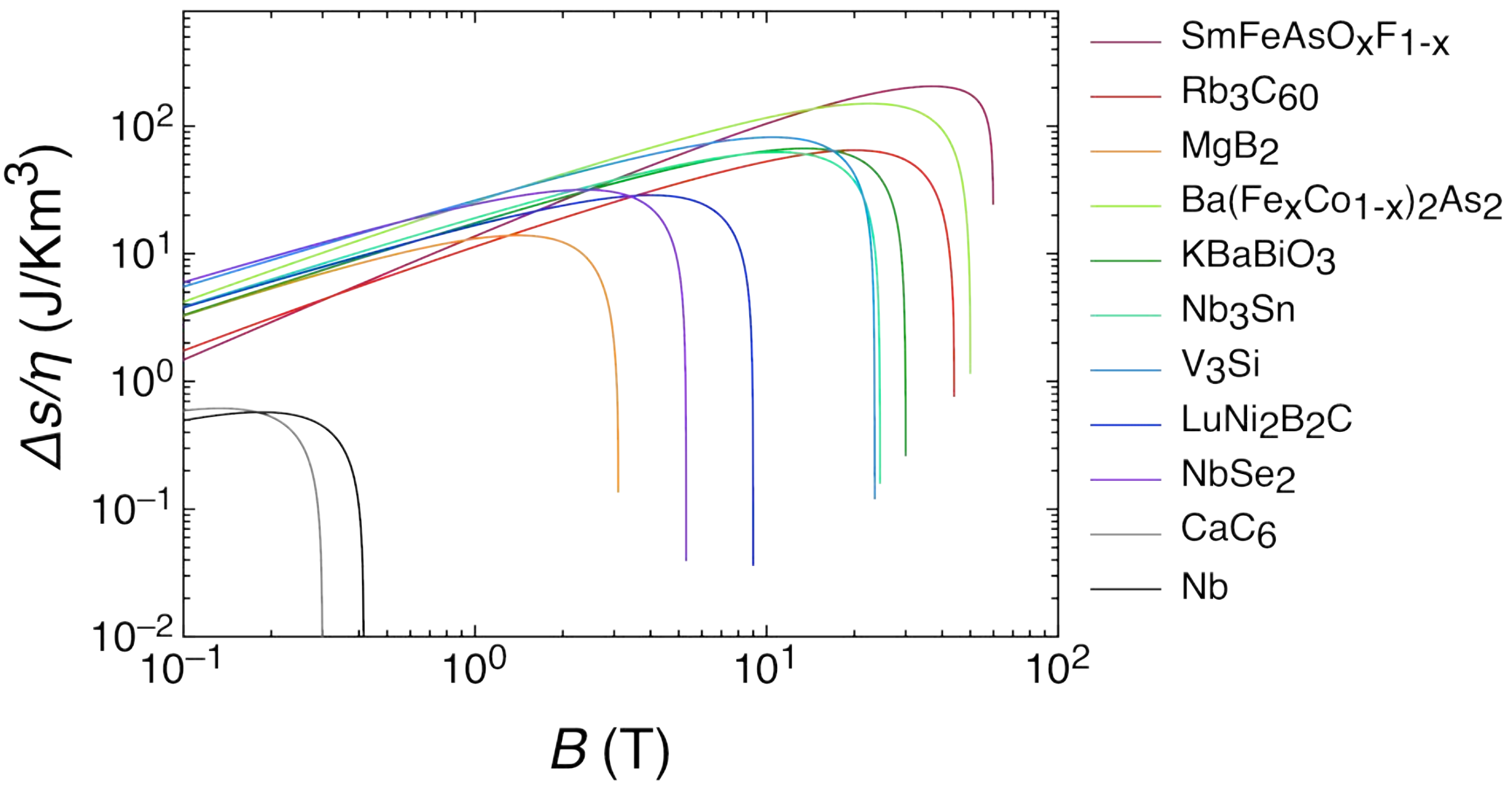

Fig. 4: Vortex-lattice melting entropies $\Delta s/\eta$ for various type-II superconductors at the $B_{\mathrm{m}}(T)$ shown in Fig. 2, a) as a function of $T$ and b) as a function of $B$.

| Substance | $T_{c0}$ [K] | $B_{c2}(0)$ [T] | $\kappa$ | $\varepsilon$ | $Gi$ | References |
|---|---|---|---|---|---|---|
| $SmFeAsO_xF_{1-x}$ | 55 | 60 | 99 | 0.125 | $1.66 \cdot 10^{-2}$ | [39] |
| $YBa_2Cu_3O_7$ | 92 | 120 | 65 | 0.125 | $4.96 \cdot 10^{-3}$ | [5, 40,41] |
| $Rb_3C_{60}$ | 29 | 44 | 90 | 1 | $6.70 \cdot 10^{-5}$ | [42] |
| $MgB_2$ | 39.0 | 3.1 | 11 | 0.097 | $4.08 \cdot 10^{-5}$ | [43,44] |
| $Ba(Fe_{1-x}Co_x)_2As_2$ | 22 | 50 | 66 | 0.666 | $2.16 \cdot 10^{-5}$ | [45] |
| $K_xBa_{1-x}BiO_3$ | 34.0 | 30 | 36 | 1 | $3.46 \cdot 10^{-6}$ | [46] |
| $Nb_3Sn$ | 18.3 | 24.5 | 34 | 1 | $9.77 \cdot 10^{-7}$ | [47,48] |
| $V_3Si$ | 16.9 | 23.5 | 22 | 1 | $1.52 \cdot 10^{-7}$ | [47,49,50] |
| $LuNi_2B_2C$ | 16.6 | 9.0 | 13 | 0.75 | $8.31 \cdot 10^{-8}$ | [51,52,53] |
| $NbSe_2$ | 7.16 | 5.3 | 11 | 0.31 | $7.88 \cdot 10^{-8}$ | [54,55] |
| $CaC_6$ | 11.5 | 0.3 | 2.1 | 0.37 | $3.35 \cdot 10^{-9}$ | [56,57] |
| Nb | 9.22 | 0.416 | 2.2 | 1 | $2.56 \cdot 10^{-10}$ | [58,59] |

Table 1: Literature values for relevant material parameters of various type-II superconductors

a)

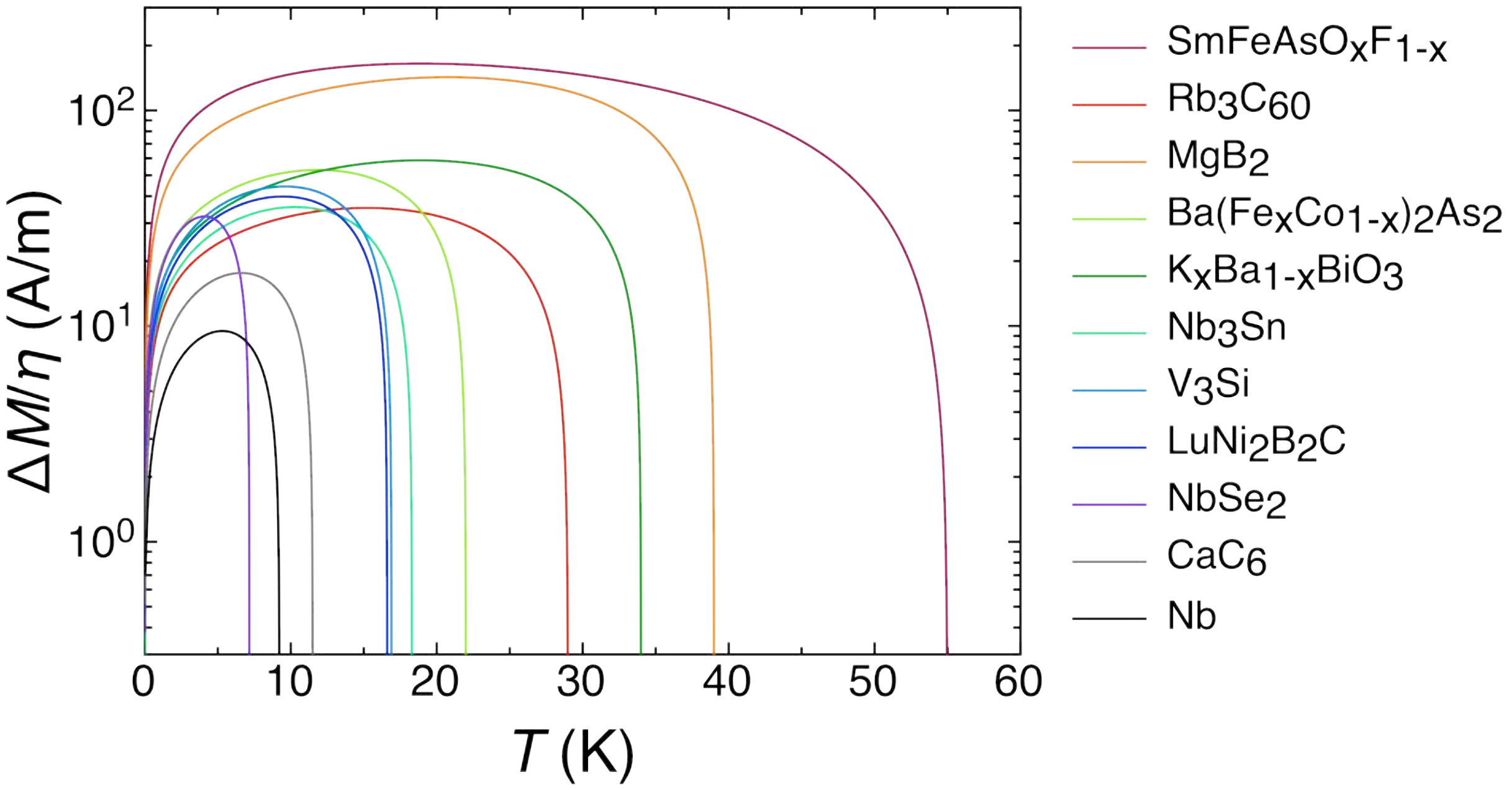

b)

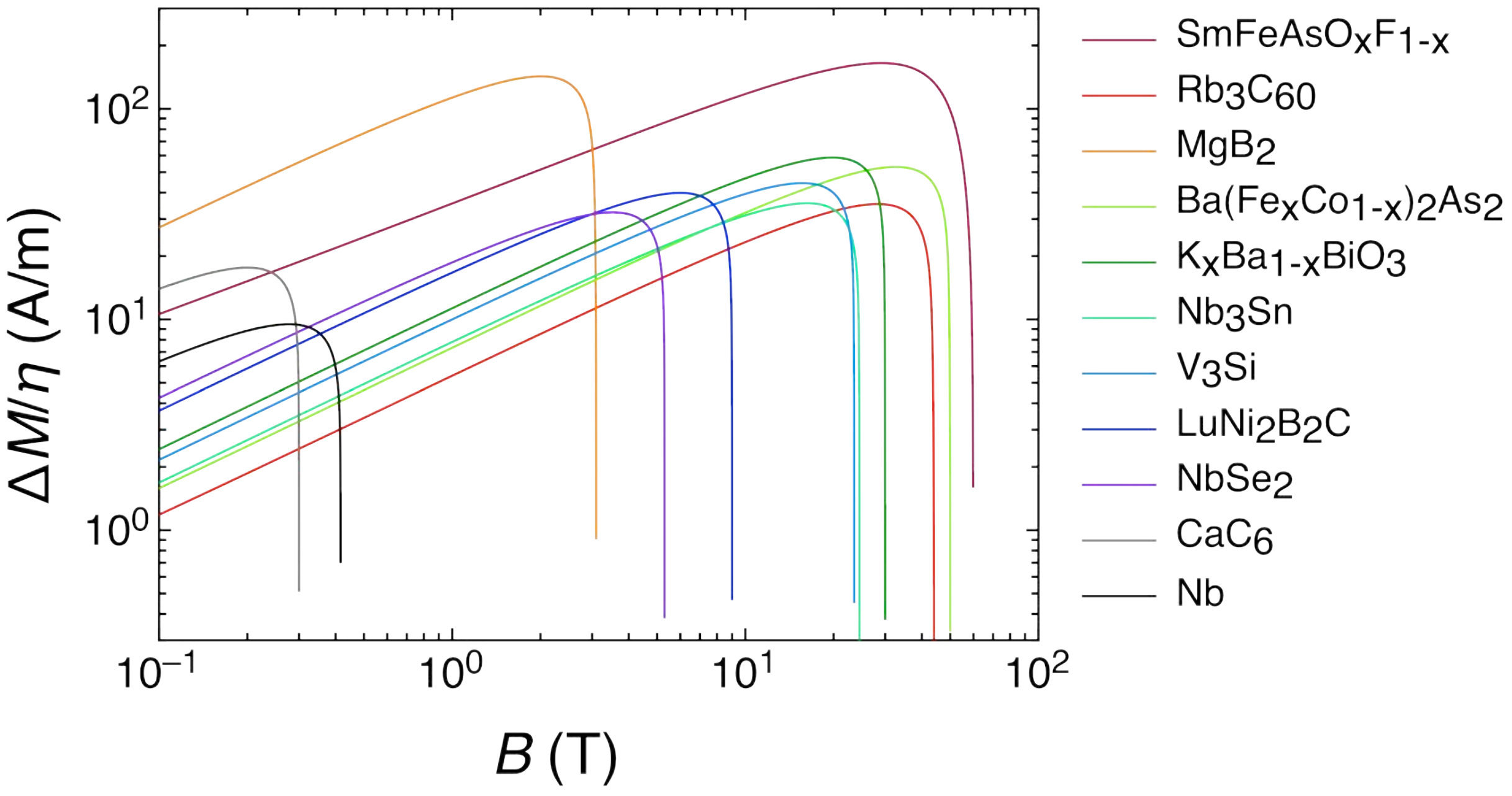

Fig. 5: Discontinuities $\Delta M/\eta$ in magnetization at the $B_m(T)$ of the materials under discussion, a) as a function of $T$ and b) as a function of $B$. To convert the given $\Delta M$ values from SI into cgs unit, the corresponding SI-values have to be multiplied with $10^{-3}$ to obtain $\Delta M$, and with $4\pi 10^{-3}$ to obtain $4\pi\,\Delta M$ in Gauss.

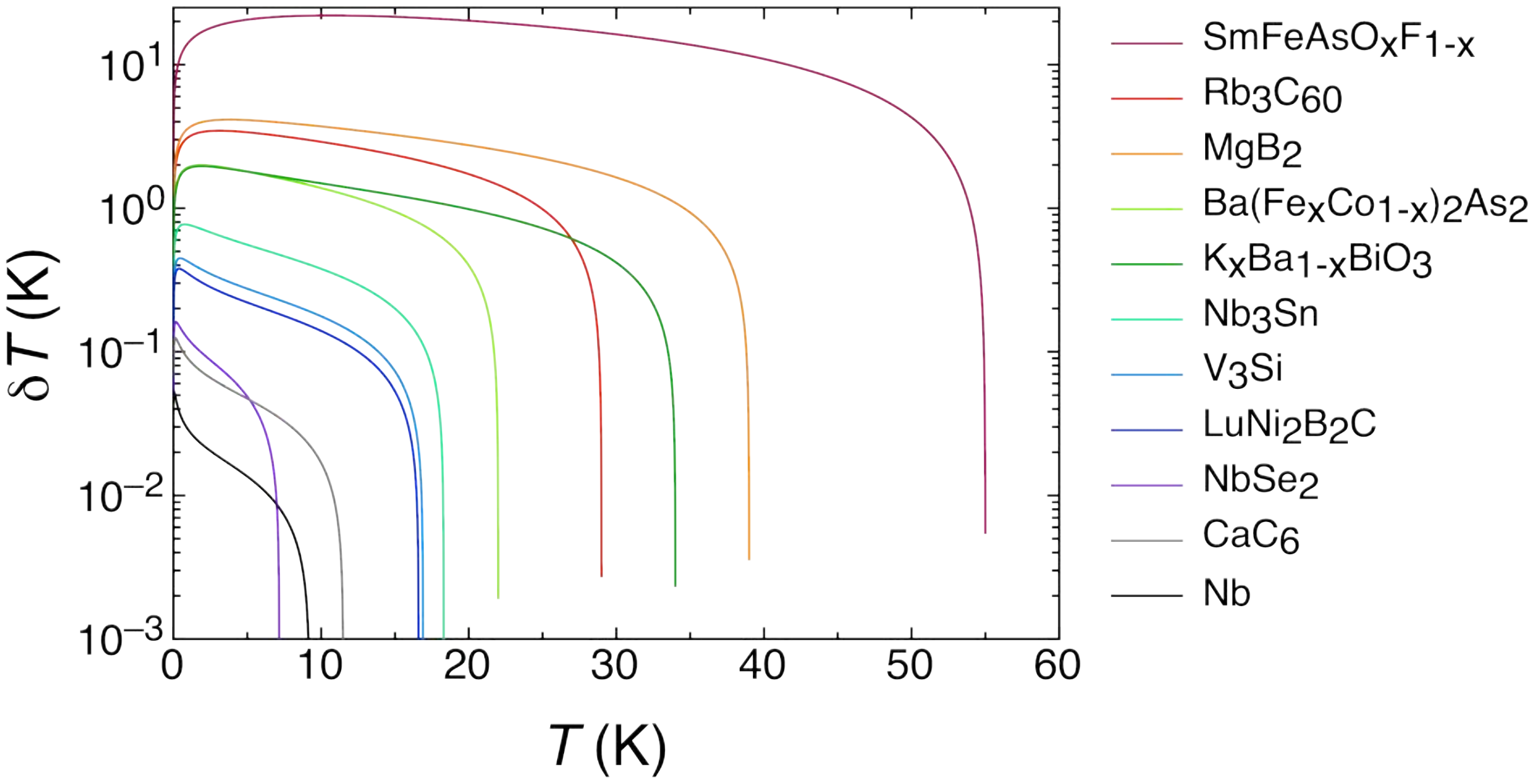

Fig. 6: Distances $\delta T$ of the melting temperatures $T_m$ from $T_{c2}$ in experiments with fixed magnetic-flux density $B$.

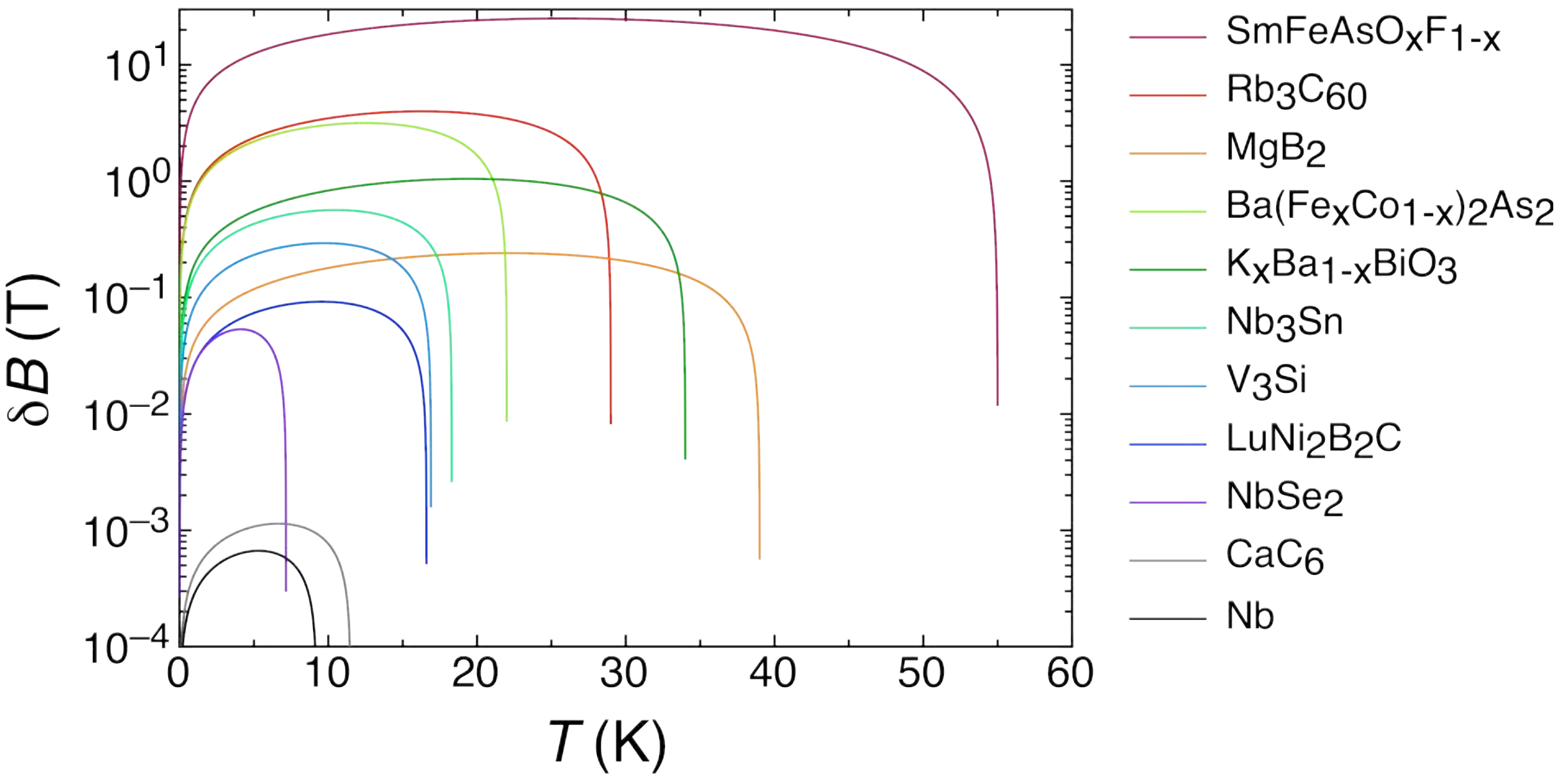

Fig. 7: Difference $\delta B = B_{c2}(T) - B_m(T)$ in experiments with constant temperature $T$.

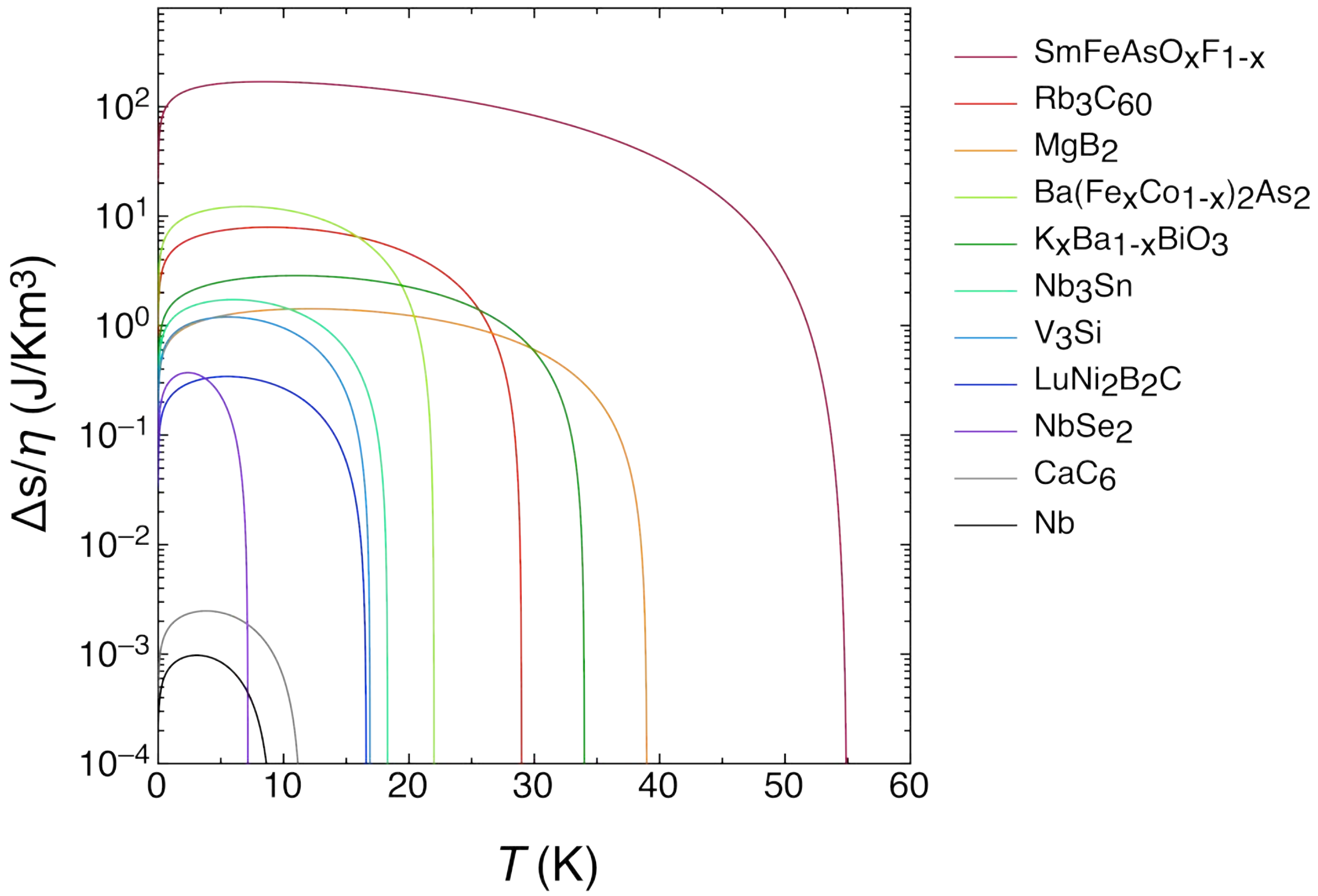

Fig. 8: Vortex-lattice melting entropies $\Delta s/\eta$ calculated as in Fig. 4a, but without the enhancement given by Eq. (15) taken into account.

## 4. Concluding discussion

As expected, the discontinuities in $\Delta s$ and $\Delta M$ are largest in compounds with a high critical temperature $T_{c0}$ and a large $Gi$. The maximum possible values for the discontinuities $\Delta s$ vary by more than two orders of magnitude between the different substances, and they are commonly attained around approximately $B \approx B_{c2}(0)/2$ (see Fig. 4b). The smallest $\Delta s$ and $\Delta M$ values can be expected in low-$T_{c0}$, low-$Gi$ materials such as Nb and $CaC_6$. Moreover, the latter materials show also low $\kappa$ values of the order of unity, and they may therefore suffer from an additional reduction of $\Delta s$ and $\Delta M$ as discussed above.

Nevertheless, even with a conservative estimate $\eta \approx 0.03$ and taking a reduction of the form $1-1/2\kappa^2$ into account, we obtain for the maximum possible $\Delta s$ in Nb $\approx 16$ mJ/Km$^3$, which should still be within reach of state-to the art calorimeters [14]. In this context we would like to mention that in Ref. [14], a marked narrowing of the expected fluctuation peak in the heat-capacity $C(T,B)$ data of Nb near $T_{c2}(B)$ has been observed, with a $C(T,B)$ that clearly exceeds the Thouless theoretical prediction [60] to explain the fluctuation contribution to the heat capacity, and notably in a field range where we find $\Delta$ to be maximum (around $B \approx 200$ mT, see Fig. 4b). A clear drop a few mK below $T_{c2}(B)$ of the diffracted intensity in corresponding small-angle neutron scattering (SANS) experiments (a quantity which is related to the magnetization $M$ [14]), might be closely related to that anomalous behaviour because both thermodynamic quantities $M$ and $C$ can be derived from the same Gibb's potential $G(T,H)$. The narrowing and enhancement of the heat capacity beyond standard expectations have been explained in Ref. [14] as possibly stemming from either fluctuation contributions that are not accounted for in Thouless' original theory, or from a coupling of the vortex system to the crystal lattice, while the sharp drop in intensity (i.e., the sudden increase in $M$ with increasing $T$) remained unexplained. It is conceivable that the observed heat-capacity contribution *in excess* to standard fluctuation theory is, along with the increase in $M$, in fact related to the missing vortex-melting entropy. A rough estimate of the excess entropy based on the $C(T,B)$ data presented in Ref. [14] is at least in line with the order of magnitude that we have calculated for $\Delta s$.

For $Nb_3Sn$, distinct peaks in $C(T,B)$ have been reported to occur in certain range of magnetic fields and for temperatures $T$ near $T_{c2}(B)$ [31], which appeared only with the aid of an additional externally applied "shaking" magnetic field. Such "shaking" techniques have been very successful to reveal the $\Delta M$ in $YBa_2Cu_3O_7$ [61] because they help the vortex lattice to come to an equilibrium state [62]. The related magnetocaloric experiments on $Nb_3Sn$ yield an estimate of $\Delta s \approx 2$ mJ/g-at K ≈ 180 J/Km$^3$ in $B = 7$ T, while the published $C(T,B)$ data suggest a $\Delta s \approx 0.3$ mJ/g-at K ≈ 27 J/Km$^3$ in $B = 6$ T [31]. Inspecting our Fig. 4b we estimate that under these conditions and with $\eta \approx 0.06$, $\Delta s \approx 4$ J/Km$^3$ at most, which is much smaller than what has been reported in Ref. [31]. We have to state, however, that $Nb_3Sn$ shows a particularly strong "peak-effect" near $T_{c2}(B)$ that becomes even more pronounced upon "vortex-shaking", with a very sharp onset as $T$ is increased towards $T_{c2}(B)$ [63]. This peak-effect is believed to be a manifestation of enhanced vortex pinning right below the upper-critical field, and the vortex lattice is therefore expected to be prone to disorder and non-equilibrium effects.

We finally want to state that the order-of magnitude estimates that we have calculated for the magnetization discontinuities, $\Delta M / \eta \approx 10$ to 150 A/m (with $\eta \approx 0.06$, $4\pi\Delta M \approx 8$ mG to 0.11 G in cgs units) are small, but should still be within the sensitivity specifications of commercial SQUID magnetometers. Nevertheless, unlike the heat capacity which is a measure of a bulk property of the vortex lattice (i.e., probing the total magnetic flux density $B$), the magnetization $M$ very is sensitive to tiny variations in $B$ because $\mu_0 M = B - \mu_0 H$ and usually $|\mu_0 M| << B$. While non-equilibrium effects due to vortex pinning may change the total flux density $B$ only slightly, they can have disastrous consequences on attempts to measure the tiny equilibrium $\Delta M < |M| << B/\mu_0$.

In the above approach to obtain reasonable estimates for the expected melting entropies $\Delta s$ and the associated discontinuities in magnetization $\Delta M$ we have deliberately ignored vortex pinning, although such an effect, if strong enough, can make a reliable measurement impossible because thermodynamic equilibrium is not reached on laboratory time scales. In addition, strong pinning makes the identification of the thermodynamic melting line $B_m(T)$ very difficult [64]. Moreover, we have not considered other possible peculiarities in the $B$-$T$ phase diagrams, such as a dimensional crossover as observed in the cuprates [65], which would affect the occurrence of a first-order melting transition of the vortex lattice as well. We have also made assumptions that should be further backed up by theory. It should in particular be clarified to what extent the enhancement of the configurational entropy $\Delta s_0$ near $B_{c2}(T)$ in Eq. (15) is really applicable in conventional superconductors. Without such an enhancement (see Fig. 8) both the expected $\Delta s$ and $\Delta M$ can be significantly smaller (by up to two orders of magnitude) than the above estimates (see Fig. 4) and may fall beyond the detection limit of an experiment. Nevertheless we believe that our considerations may serve as guideline for searching signatures of a vortex-lattice melting transition in conventional and novel superconductors, and for estimating the correct order of magnitude to be expected for the related thermodynamic quantities.

## Acknowledgments

We thank to V. B. Geshkenbein and S. Weyeneth for valuable discussions. This work was supported by the Schweizerische Nationalfonds zur Förderung der wissenschaftlichen Forschung, Grant. No. 20-131899.

## Figure Captions

Fig. 1: Melting lines (according to Ref. [35] with $c_L$ = 0.20) for different values of *Gi*.

Fig. 2: Melting entropy $\Delta s$ and melting line $B_m(T)$ of $YBa_2Cu_3O_7$, calculated from the parameters given in table 1 and Eqs. 1, 5, 12 and 15. The $\Delta_.$ -data fit best for $\eta$ = 0.077.

Fig. 3: Expected melting lines $B_m(T)$ for various type-II superconductors, calculated from Eqs. (1) and (5) with $c_L$ = 0.20 and the material parameters given in table 1 (solid lines). The $B_{c2}(T)$-lines (dotted lines) have been plotted only for $SmFeAsO_xF_{1-x}$, $Rb_3C_{60}$, $MgB_2$, and $Ba(Fe_xCo_{1-x})_2As_2$ for which they appear distinct from $B_m(T)$ in this representation.

Fig. 4: Vortex-lattice melting entropies $\Delta s/\eta$ for various type-II superconductors at the $B_m(T)$ shown in Fig. 2, a) as a function of *T* and b) as a function of *B*.

Fig. 5: Discontinuities $\Delta M/\eta$ in magnetization at the $B_m(T)$ of the materials under discussion, a) as a function of *T* and b) as a function of *B*. . To convert the given $\Delta M$ values from SI into cgs unit, the corresponding SI-values have to be multiplied with $10^{-3}$ to obtain $\Delta M$, and with $4\pi 10^{-3}$ to obtain $4\pi\,\Delta M$ in Gauss.

Fig. 6: Distances $\delta T$ of the melting temperatures $T_m$ from $T_{c2}$ in experiments with fixed magnetic-flux density *B*.

Fig. 7: Difference $\delta B = B_{c2}(T) - B_m(T)$ in experiments with constant temperature *T*.

Fig. 8: Vortex-lattice melting entropies $\Delta s/\eta$ calculated as in Fig. 4a, but without the enhancement given by Eq. (15) taken into account.